# Turbulent velocity measurements in high Reynolds cryogenic helium facilities at Service des Basses Températures (SBT)

S Kharche[1], J P Moro[2], C Baudet[3], B Rousset[1], A Fuchs[4], J Peinke[4], A Girard[1]

[1] Université Grenoble Alpes, CEA INAC-SBT, F-38000 Grenoble, France
[2] CEA-DEN-DANS-STMF, F-38000 Grenoble, France
[3] Université Grenoble Alpes, LEGI, F-38041 Grenoble, France
[4] Carl von Ossietzy Universitaet Oldenburg, Germany

Corresponding author: alain.girard@cea.fr

**Abstract**. Due to its very low viscosity, cryogenic helium has been used for years to generate high Reynolds turbulent flows. The measurement of velocity fluctuations in such high Reynolds flows is however a challenging issue, as it is necessary to develop small size (typically micron-length), robust sensors, to measure the whole spectrum of fluctuations of the velocity, which may extend to hundreds of kHz and possibly higher. SBT has developed for years different facilities, in particular taking benefit of the refrigeration capacities available at CEA Grenoble. In this article we present the current status of developments of hot wire sensors at CEA. Different characterizations of Wollaston hot wires are shown, and measurements of velocity fluctuations in different conditions, in normal helium as well as in superfluid helium are shown.

## 1. Introduction

Turbulence has a very broad spectrum of occurrences in nature, industry and every-day-life. Indeed, geophysical flows (stratospheric, oceanic, etc) are always turbulent, as they reach very high Reynolds numbers. Mixing is enhanced through turbulence, making easier e.g. combustion in multifluidic flows of engines. The efficiency of heat exchangers, and pressure drops in pipes, are deeply influenced by turbulent heat/momentum transfer. Turbulence also plays an important role to generate energy using wind turbines, and it also decides the gas consumption of your car. Although much progress has been performed in the characterization and modelling of turbulence in the past years, there is still a strong need of understanding the mechanisms of turbulence. The most dramatic step forward in understanding turbulent flows occurred through the celebrated work by Kolmogorov [1,2] based on a statistical analysis of turbulence at infinite Reynolds number. According to Kolmogorov's first description of the cascade of energy along the different scales [1], there is a domain where this cascade is self-similar (i.e. the transfer of energy from larger to smaller scale is independent on the scale). This domain, where viscosity plays no role, is called the inertial range, intermediate between the forcing (or integral) scale, and the dissipative scale, where viscosity dissipates the energy. It is therefore of high interest to generate laboratory flows with high Reynolds number and large inertial range, in order to compare theoretical and experimental results with a great accuracy. Moreover, it is of great interest to achieve laboratory flows of similar Reynolds number as geophysical flows. These are the main reasons to use cryogenic helium to generate turbulent flows. Indeed, helium is the fluid with the lowest kinematic viscosity, ν (see table 1), so that large Reynolds number $R_e \equiv \frac{U\,L}{\nu}$ can be achieved on a laboratory scale L with a characteristic velocity U. However, large Reynolds numbers induce a small dissipative (also called Kolmogorov) scale $\eta$, where energy is dissipated through viscous effects. Indeed, one can show that $L/\eta \sim R_e^{3/4}$. As a consequence, interesting measurements should provide a space resolution of the order of dissipative scale, typically of a few µm. The quest for small sensors has started in the 1990s, and is still ongoing [3-5]. Many small size sensors were achieved, but so far with little reliability. We present here a robust sensor, which has been developed at CEA, whose length is significantly larger than the Kolmogorov length, but provides however much information on the inertial range. In this paper, we describe the sensor, how it is operated, and show some results in different experimental conditions.

**Table 1.** Kinematic viscosity, ν of some elements. GHe: gaseous helium; LHe: liquid helium.

|  | Temperature (K) | Pressure (bar) | Kinematic Viscosity (m² s⁻¹) |
|---|---|---|---|
| **Air** | 298 | 1 | 1.6 10⁻⁵ |
| **Water** | 298 | 1 | 0.88 10⁻⁶ |
| **SF₆** | 273 | 10 | 1.8 10⁻⁷ |
| **GHe** | 4.5 | 1 | 0.9 10⁻⁷ |
| **LHe** | 2.25 | 3 | 2.1 10⁻⁸ |

## 2. Wollaston hot wires: fabrication, characterization and operation.

Hot wire anemometry is a well-known technique for characterization of the turbulent velocity field [6,7]. It still provides unsurpassed quality results in terms of signal to noise ratio, resolution, and frequency response. Moreover, optical diagnostics (for Laser Doppler Velocimetry, Particle Image Velocimetry or Particle Tracking Velocimetry) are still not standard in cryogenic experiments, and are so far not available in our facilities.

*2.1. Fabrication*

No hotwire of length in the range of a few hundreds of µm is available commercially. Therefore, they are built at CEA according to a patented process [8]. The base material is a Wollaston wire (Platinum-Rhodium wire in a Silver coating of 50 µm in diameter). The diameter of the PtRh wire is either 1.3 µm, or 0.65 µm for the shortest (and most fragile). The silver coating is chemically etched over a small distance (so far between 200 and 400 µm). Then, in order to avoid the breaking of the wire during the cool down process, the wire is slightly bent.

*2.2. Operating principle*

Hot wire anemometry operates according to the "wind chill" principle, more precisely it consists in measuring the heat transfer from a wire to the ambient moving fluid. Consider a wire of mass m, diameter $d$ and length $l$, of heat capacity $C$, temperature $T_w$ and resistance $R(T_w)=R_w$, heated by a current $I$, in a fluid of temperature $T_a$, thermal conductivity $k_f$, viscosity $\nu_f$, moving with a velocity $U$ with respect to the wire. Neglecting the thermal conduction at the ends of the wire ($l>>d$, [6]), neglecting radiative transfer, the power balance equation of the wire is:

$$m\,C\frac{dT_w}{dt} = R_w\,I^2 - \pi\,l\,k_f\,(T_w - T_a)\,Nu \qquad (1)$$

using the Nusselt number, which describes the efficiency of the heat transfer with respect to pure conduction. King [9] proposed a square root law for the Nusselt versus Reynolds number. A generalized King's law can be written (defining the Reynolds number at the wire: $Re_w \equiv U\,d/\nu_f$):

$$Nu = a + b\,Re_w^n \qquad (2)$$

The variation of the resistance of the wire with temperature is essential for hot wire anemometry. Usually it is assumed that the resistance of the wire depends linearly on temperature. This is because the slight change in the temperature or resistance of the hot wire can be measured as a result of small change in convective losses encountered by hot wire. In our case, the dependence of the resistance with temperature is more complicated (resistance is almost constant from 0 to 20 K, then it slightly increases with temperature, till it reaches a quasi linear dependence with temperature above 80 K. For simplicity, we will assume a linear dependence of resistance with temperature within a range of 60 to 90 K.

$$R_w = R_0(1 + \alpha (T_w - T_a))  \quad (3)$$

defining: $A = \frac{\pi\, l\, k_f}{\alpha R_0} a$ and $B = \frac{\pi\, l\, k_f}{\alpha R_0} \left(\frac{d}{v}\right)^n b$, we obtain the following differential equation for $R_w$: (see for instance [7]):

$$\frac{m\,C}{\alpha R_0\,(A - I^2 + BU^n)} \frac{dR_w}{dt} + R_w = \frac{A + BU^n}{(A - I^2 + BU^n)} R_0 \quad (4)$$

From this equation we can derive two conclusions:
(i) In steady state (or neglecting the time derivative), and multiplying by the current I in the wire, we obtain the relation between the measured voltage of the hot wire and the flow velocity, which provides the calibration of the measurement;
(ii) Associated with equation (4), the sensor has a time constant which is given by:

$$\tau = \frac{m\,C}{\alpha R_0\,(A - I^2 + BU^n)} \quad (5)$$

which is independent on *l*. From this, we conclude that a hot wire provides a measurement of the velocity of the flow after some calibration, and that it has a certain time constant, allowing to measure fluctuations up to a certain frequency. There are usually different types of measurements with hot wires: when the current is taken constant, this is called CCA operation (Constant Current Anemometry); it may be preferred to work at constant temperature (which requires some electronics and control loop – in that case the measurement is provided by the current injected in the loop) which is called CTA (Constant Temperature Anemometry). CTA was used in [10]. Constant Voltage Anemometry (CVA) is also possible. Advantages and drawbacks of the techniques are summarized in [6-7]. In what follows we will describe CCA operation.

*2.3. Characterization: calibration and time constant.*
Calibration and time constant measurements are consequently of primary importance to use the experimental data from hot wires. If we consider small perturbations in current or velocity with respect to steady state quantities, equation (4) can be linearized, and we obtain a very similar equation where the coefficient are functions of the mean values which are independent of time. The time constant for a variation of current, or a variation of velocity, is the same, and is given by equation (5) – using mean quantities. Hence, the sensor is a first order filter, with time constant given by equation (5), and this time constant can be measured by applying a small fast incremental step of current. This is what is done below to determine τ. The procedure for calibration of the velocity in SHREK is also given.

## 3. Experiments

*3.1. Superfluid High REynolds von Kármán experiment (SHREK).*
The SHREK experiment has been developed to study High Reynolds turbulence in Cryogenic Helium. It is a large von Kármán flow which can use normal cryogenic helium as well as superfluid helium, which is of additional interest for fundamental turbulence research. The in-depth details of this experiment can be found in [11]. In order to calibrate the velocity sensor, we generate a quasi solid flow by turning the two propellers at the same velocity (up to 1 Hz in our case). Different hot wires are installed at different places (at 40 mm from the wall, positioned to measure the azimuthal velocity of the flow). The flow driven by such a co rotation is a low turbulence level solid flow moving at the velocity close to the velocity of the turbines. By changing the rotation speed of propellers, we are able to generate a calibration curve (figure 1). This procedure has been applied both in normal and superfluid helium. Concerning the determination of time constant, a small pulse of current (< 10% of the total applied current) is suddenly injected, and the response of the hot wire is analysed: we can see that it takes some time for the voltage to reach its stationary value, and this time is a direct measurement of the time constant. Figure 2 presents the time constant as a function of the velocity.

*3.2. Experiments in normal helium.*

Figure 1 shows a calibration curve, together with a fit assuming King's law (i.e. square root dependence of the Nusselt with the velocity). This figure shows that King's law (curve in orange) provides a satisfactory description of the calibration. Concerning the time constant (figure 2), we can verify that it depends on the velocity, as predicted in equation (5). A fit is presented, which is compatible with equation (5), although the coefficients are somewhat different from what is expected from theoretical values in (5). However, the general shape of the time constant variation with velocity is satisfactory.

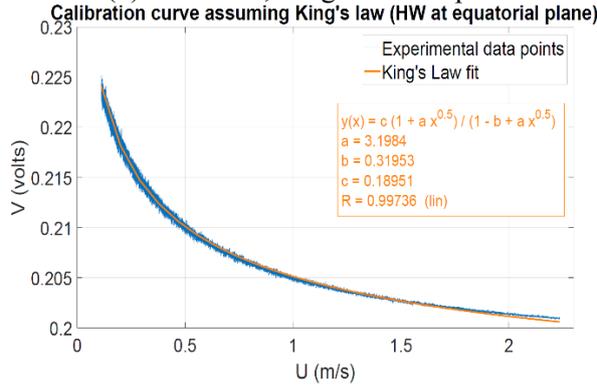
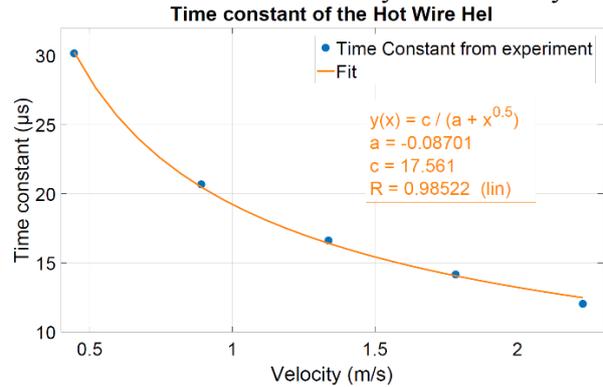

Figure 1. Calibration curve in SHREK and King's law fit.    Figure 2. Evaluation of the time constant versus velocity.

*3.3. Experiments in superfluid helium (HeII).*

The same experiments (calibration, time constant measurement) were performed in superfluid helium (T=2 K). As the power balance in HeII is different from the situation in normal helium, the temperature of the hot wire was lower in HeII than in normal helium (HeI) for the same current. Therefore, we chose to increase the current in the hot wire compared with what happened in HeI (13 mA instead of 8 mA). The calibration curve (figure 3) and time constants (figure 4) were then obtained.

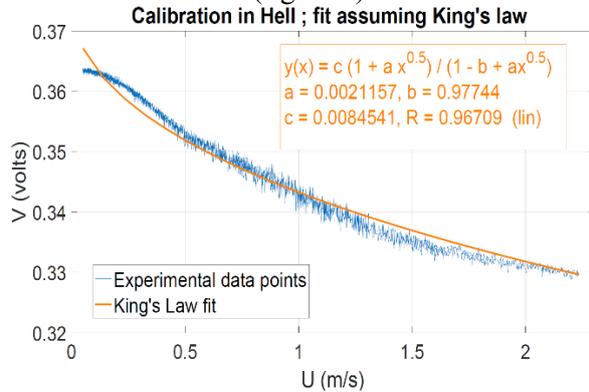
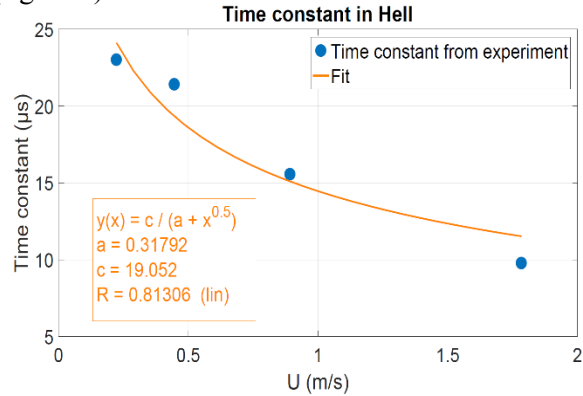

Figure 3. Calibration curve in HeII    Figure 4. Time constant in superfluid He and fit.

The fit with a King's law is no longer satisfactory, as the heat transfer mechanism is different in superfluid than the one explained in equations (1)-(2), which is in agreement with [10]. However, the sensor is also sensitive to the velocity, as in normal helium; but it is necessary to have a stronger velocity to influence the voltage, compared to HeI. Concerning time constants, the shape is also changed compared to HeI: it seems that the time constant is less sensitive to the velocity than in HeI.

In superfluid helium the spectra at zero (or very small) velocity are totally different from those in HeI. Indeed, a strong energy content is present, up to very high frequencies (this was already found in [10]). The overheating affects also the shape of the spectra: figure 5 shows two spectra in HeII, for two heating currents. These spectra were obtained in the Hejet experiment [12], at a temperature of 2 K. Similar spectra are obtained in SHREK at zero velocity. This occurrence of high energy content at high frequencies is also encountered when the SHREK experiment is run at its highest turbulence level, but the difference with turbulence spectra in normal helium becomes less and less pronounced. Figure 6 hows two spectra in HeI and HeII, obtained in the same conditions (rotation of propellers): only the temperature changes from normal to superfluid. Both spectra show the well-known -5/3 slope over the inertial range as predicted by Kolmogorov [1]. Surprisingly, HeII seems to offer a longer -5/3 dependence (i.e. towards higher frequencies) than HeI.

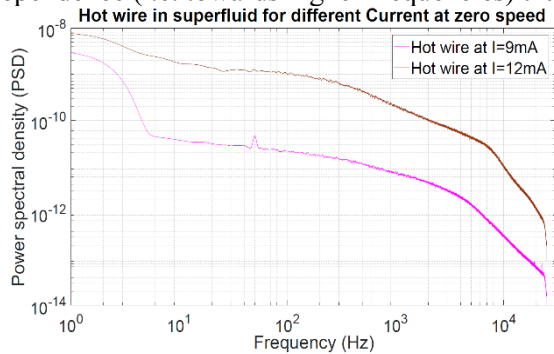
Figure 5. Spectrum at zero velocity in HeII for two different heating currents.

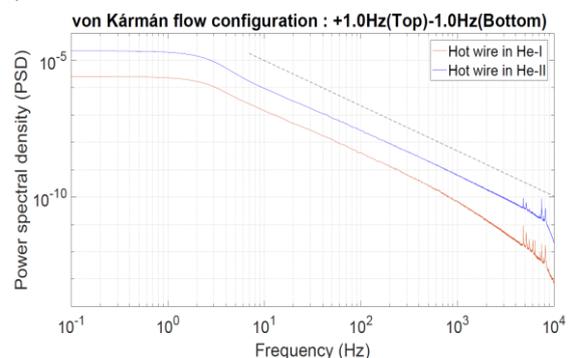
Figure 6. Two spectra (He I / HeII) obtained in the same experimental conditions (high turbulence level).

## 4. Discussion and prospects.

From these studies we have verified that hot wires (usually working at high temperature in a room temperature fluid) can be used at cryogenic temperature. In HeI, the operation of such Wollaston hot wires only requires some bending of the etched part, in order to avoid breaking of the sensor during cool down. In HeII we confirm that the hot wire can measure the velocity, but the calibration curve is different from the usual King's law. This is due to the totally different heat exchange process in superfluid helium, which evacuates energy more efficiently than in normal helium. In both cases however, the hot wire itself is surrounded by a thin (µm range) layer of normal helium. At high turbulence level however, the cascade process in velocity space seem to be the same in normal and superfluid helium in the inertial range, up to a certain limit (in our case of order of a few kHz) beyond which the high energy content in HeII is no longer related to the usual cascade process. In this study, we have also determined the time constants of the hot wires, which are of order of 20 µs. Our hot wire behaves as a low pass filter of frequency $1/2\pi\tau \sim 8$ kHz. This good high frequency response is connected with the small diameter. In CTA and CVA operation, a higher cutoff frequency could be obtained, but this requires additional electronics. In HeI we saw that the spectrum decays faster than -5/3 beyond 3 kHz (figure 6), which can be explained by the length of the sensor [7], which is much larger than the Kolmogorov length. Therefore we continue the development of hot wires, in order to approach the Kolmogorov (dissipative) length. Calibration of hot wires in SHREK is based on the assumption of solid rotation. We have developed the Hecal cryostat to provide a reliable tool for calibration of hot wires, which will soon start its operation.

**Acknowledgements.**
We acknowledge the support of the EUHIT program funded in the FP7 program, (Grant Agreement No. 312778). We also acknowledge the support of the LANEF program (ANR-10-LABX-51-01).